\documentclass[preprint,12pt, a4paper]{elsarticle}

\usepackage{amssymb}

\usepackage{lineno}
\usepackage{hyperref}

\usepackage{float}
\restylefloat{table}

\usepackage{color}

\journal{SoftwareX}

\begin{document}

\begin{frontmatter}

\title{Adapting LIGO workflows to run in the Open Science Grid}


\author[UCSD]{Edgar Fajardo}
\author[UCSD]{Frank Wuerthwein} 
\author[Mortgridge]{Brian Bockelman}
\author[Mortgridge]{Miron Livny}
\author{Greg Thain}
\author[GATech]{James Alexander Clark}
\author[LIGO]{Peter Couvares}
\author[LIGO]{Josh Willis}

\address[UCSD]{University of California San Diego, 9500 Gilman Dr, La Jolla, CA 92093}
\address[Mortgridge]{Mortgridge Institute, 330 N Orchard St, Madison, WI 53715}
\address[GATech]{School of Physics, Georgia Institute of Technology, Atlanta, GA 30332, USA} \address[LIGO]{LIGO, California Institute of Technology, Pasadena, CA 91125, USA}

\begin{abstract}
During the first observation run the LIGO collaboration needed to offload some of its most, intense CPU workflows from its dedicated computing sites to opportunistic resources. Open Science Grid enabled LIGO to run PyCbC, RIFT and Bayeswave workflows to seamlessly run in a combination of owned and opportunistic resources. One of the challenges is enabling the workflows to use several heterogeneous resources in a coordinated and effective way.

\end{abstract}


\end{frontmatter}

\section*{Required Metadata}
\label{MetadataSec}

\section*{Current code version}
\label{CurrentCodeSec}

\begin{table}[H]
\begin{tabular}{|l|p{6.5cm}|p{6.5cm}|}
\hline
\textbf{Nr.} & \textbf{Code metadata description} & \textbf{Please fill in this column} \\
\hline
C1 & Current code version & v3.6.1 \\
\hline
C2 & Permanent link to code/repository for this code version & \href{url}{github.com/glideinWMS/glideinwms} \\
\hline
C4 & Legal Code License   & BSD License \\
\hline
C5 & Code versioning system used & git \\
\hline
C6 & Software code languages, tools, and services used & Python2 \\
\hline
C7 & Compilation requirements, operating environments \& dependencies & RedHat6 or RedHat7\\
\hline
C8 & Link to developer documentation/manual & \href{url}{glideinwms.fnal.gov} \\
\hline
C9 & Support email for questions & \href{url}{glideinwms-support@fnal.gov} \\
\hline
\end{tabular}
\caption{Code metadata (mandatory)}
\label{MetadataTable} 
\end{table}


\section{Motivation and significance}

In order to reach the scientific and discovery goals of the LIGO Collaboration several pipelines of CPU intensive workflows are run. During certain times the pipelines compete for computing resources at the LIGO-owned computing laboratories. There is opportunity to migrate some of these pipelines from dedicated resources to a combination of owned and opportunistic resources.

The LIGO collaboration worked with the Open Science Grid (OSG)\cite{osg} to enable PyCBC\cite{PyCbCSoft}, RIFT\cite{rift} and Bayeswave \cite{bayeswaves1}\cite{bayeswaves2} workflows to run on the grid. These workflows share the same structure. They are made of several thousand individual tasks or jobs, which require no communication between them. The task runtime is in the order of hours. This intrinsically parallel formulation made them a candidate for the Distributed High Throughput Computing (DHTC) model in OSG. The distributed model generates a data distribution challenge: the LIGO experiment data is produced at the interferometer locations and then stored at a few computing centers. The problem lies then in distributing the  data to all the participating computing centers around the world for the workflows to consume. The OSG solution to this data delivery problem is through Stashcache \cite{stashcache}. In a nutshell, Stashcache is a file block caching technology based on XRootD \cite{dorigo2005xrootd} that can deliver on-demand high volumes of data to the jobs. These jobs use GeoIP to retrieve data of the nearest cache from a set of caches conveniently located around the world.

\section{Software description}
The DHTC model in OSG is powered by the Glidein Workload Management System (GlideinWMS) system\cite{glideinwms}\cite{GlideinWMSSoft}. It is a pilot model system in which resources at heterogeneous sites are gathered and presented to the scientist as one single homogeneous pool of resources. GlideinWMS is based on the HTCondor\cite{htcondor_soft} batch system and it is designed to create a changing pool of resources based on demand.

\subsection{Software Architecture}
\subsubsection{HTCondor Architecture}
\begin{figure}[h]
\centering
\includegraphics[width=12cm,clip]{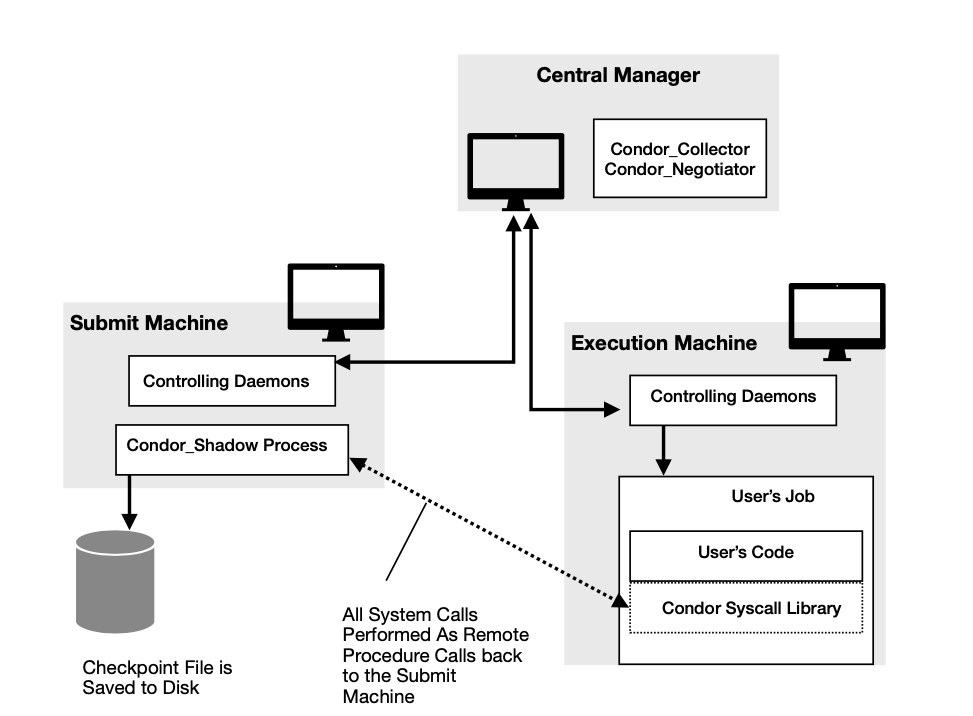}
\caption{Description of HTCondor Architecture \cite{HTCondorDoc}.}
\label{fig-HTCondorArch}       
\end{figure}

The HTCondor batch system architecture is made of three components: scheduler (schedd), central manager and computing machines (see Figure \ref{fig-HTCondorArch}). The scheduler is the multi user client facing part of the architecture and takes care of submitting the jobs, maintaining the job queue, and transferring the input and output files needed for the job from itself to the compute nodes. In a usual set up (including the one used by LIGO collaboration) several schedd are deployed to serve a single pool for scalability and fault tolerance reasons.

The second component of the HTCondor architecture is the central manager. The central manager is made of two daemons: the collector and the negotiator. The collector daemon tracks all the information on the pool. This includes which computing machines are busy/idle and which users have idle jobs in the queues (schedds). The negotiator uses the information from the collector to decide which job matches which resources and among users who have the best priority to use those resources.

The final part of the HTCondor Architecture is the daemon that runs in the computing nodes called StartD. The StartD daemon runs on every compute node in an HTCondor pool and informs the collector of the machine usage, when idle, when busy. Once it is assigned a job it contacts the corresponding schedd to start the job.

\subsubsection{GlideinWMS Architecture}

\begin{figure}[h]
\centering
\includegraphics[width=12cm,clip]{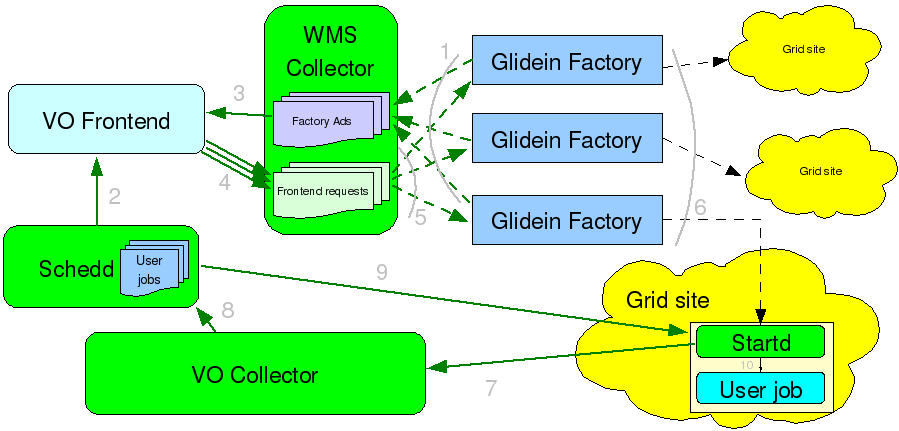}
\caption{Description of GlideinWMS architecture \cite{GWMSDoc}.}
\label{fig-gwmsArch}       
\end{figure}

The GlideinWMS pilot system builds on top of the HTCondor architecture (see Figure \ref{fig-gwmsArch}) and introduces two more pieces: factories and frontends. The latter is a set of Python daemons that continuously query the submit hosts (step 2 in Figure \ref{fig-gwmsArch}) in a single HTCondor pool and calculate the demand for resources. Based on this demand, the frontend asks the factory to submit pilots on its behalf to a grid site (step 5 in Figure \ref{fig-gwmsArch}).

The factory submits the pilots  based on the pressure requested by the frontend (step 6 in Figure \ref{fig-gwmsArch}). In order for the factory to submit to a site the site must have a Compute Element (CE) which ``translates" grid submissions into local batch system submissions. The frontend securely sends its credentials to the factory to be presented to the CE on the frontend's behalf. Once a pilot is submitted and is running at a site batch system it contacts a web server running in the factory to download configurations and the HTCondor binaries, then the pilot downloads frontend specific configurations from the frontend's web server. Finally it starts the HTCondor Daemons (as an unprivileged user) and connects back to the pool collector (step 7 in Figure \ref{fig-gwmsArch}). From this point on a pilot looks just like any other resource in an HTCondor pool. The StartD advertises its capabilities to the Collector (step 8 in Figure \ref{fig-gwmsArch}) and the schedd starts a job from its queue into the pilot (step 9 in Figure \ref{fig-gwmsArch}.

The recommend usage is that each frontend is managed by a scientific community which we will call from now on a Virtual Organization (VO). The factory(s) are centrally operated services that can serve multiple organizations and hence reduce operational costs \cite{factOps}.

\subsection{Software Functionalities}
A GlideinWMS pool can gather resources from several heterogeneous grid sites. The  type of CE varies the most among sites. The Factory makes extensive use of Condor-G \cite{htcondor_soft} capabilities to submit to CREAM\cite{CREAMCE}, ARC-CE \cite{NorduGrid} and HTCondor-CE \cite{HTCondorCE} as well as to several commercial cloud providers like Amazon Web Services and Google Cloud. 

The strength of this architecture lies in both the breadth and scale of the resources that can be gathered. 
The scalability of a GlideinWMS pools has been measured to exceed more than $200k$ running jobs \cite{CondorLimits}. GlideinWMS handles mixed sets of GPU and CPU workloads efficiently among heterogeneous resources \cite{OSGGPu}. Moreover the schedds can move several GigaBytes/sec of traffic to and from the compute nodes \cite{HTCondorTransfers}. Finally GlideinWMS, and HTCondor incorporate the ability for each individual user to run tasks in the container environment of their choice via integration with Singularity \cite{singularity_software}\cite{singularity_cvmfs}.

\section{Illustrative Examples}
\begin{figure}[h]
\centering
\includegraphics[width=12cm,clip]{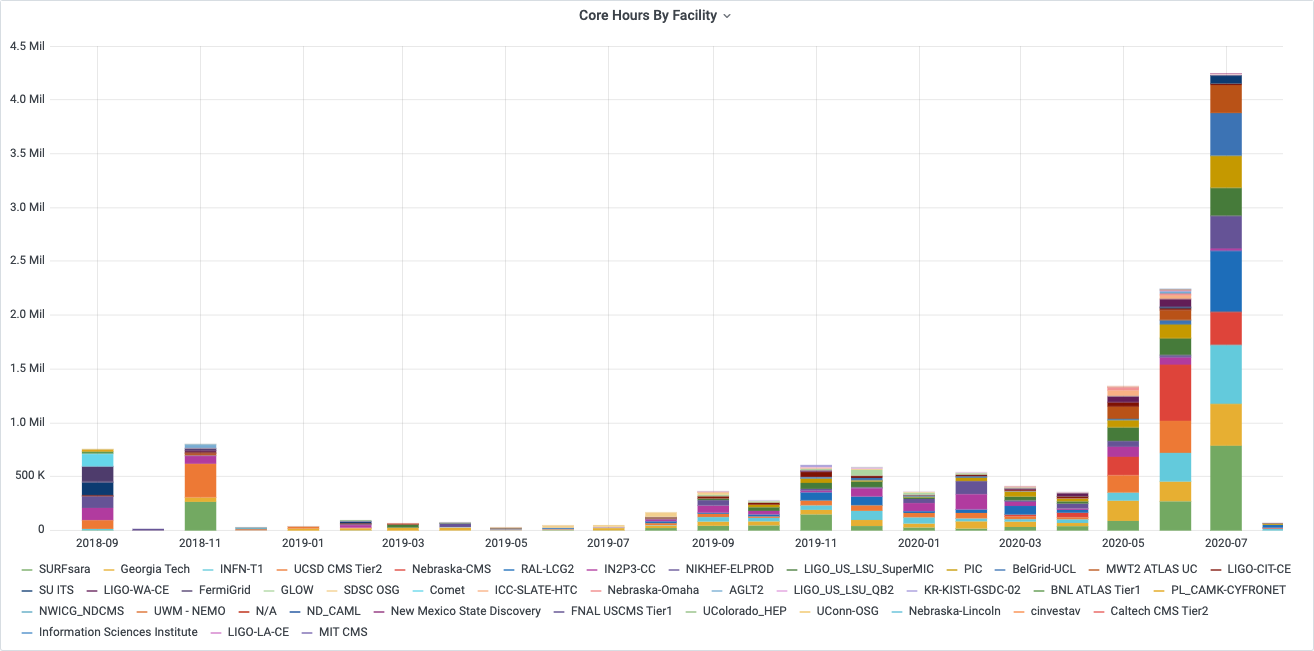}
\caption{Site usage distribution of LIGO on OSG for the last two years. Vertical axis is CPU core hours per month, peaking at 4.5 million hours per month, or an average of roughly 6,000 cores.}
\label{fig-corehours}       
\end{figure}

These capabilities are exercised by LIGO. It uses VIRGO resources on WLCG \cite{wlcg} thanks to the factory's ability to submit to different types of CEs. Moreover it uses several HPC sites, like Comet \cite{Comet} and BlueWaters\cite{BlueWaters}, in addition to traditional resources in OSG. Figure \ref{fig-corehours} shows how LIGO can consume resources from several sites that are very heterogeneous and need not communicate among themselves to collaborate.

\begin{figure}[h]
\centering
\includegraphics[width=12cm,clip]{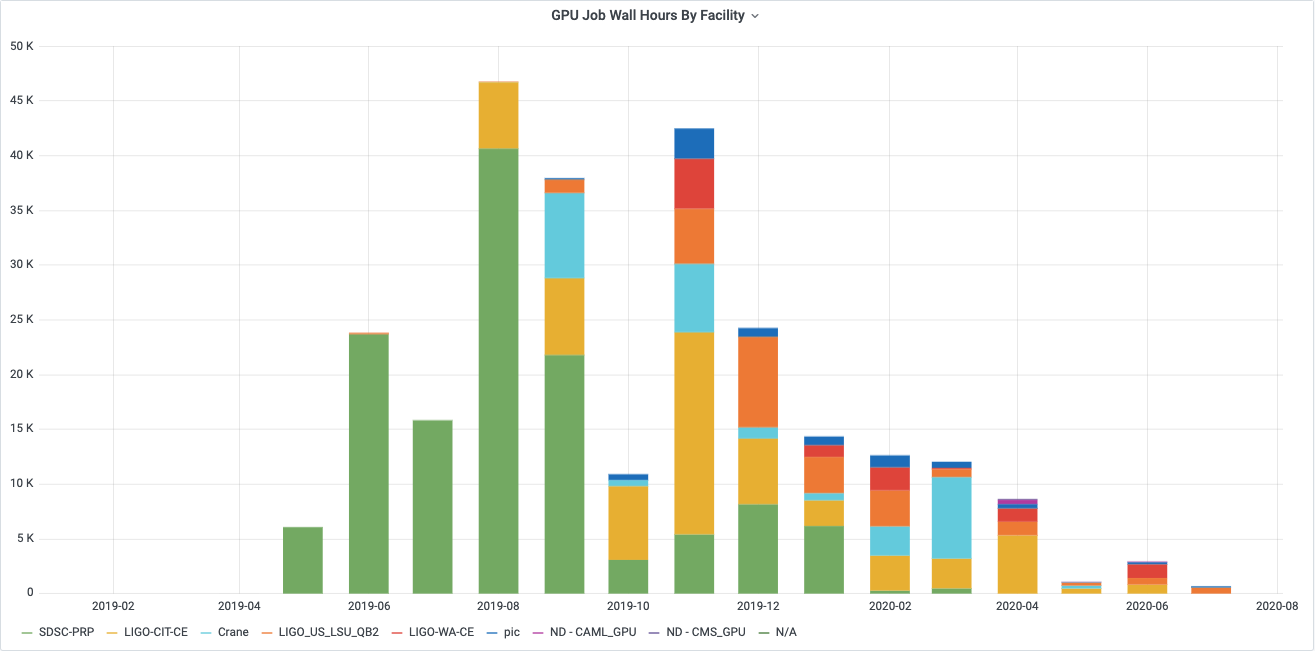}
\caption{Site usage distribution of LIGO GPU usage for the last year. Vertical axis is GPU hours per month peaking at 45,000, or an average of roughly 60 GPUs.}
\label{fig-gpuhours}       
\end{figure}

Since the RIFT pipeline was adapted to run in a distributed enviroment GlideinWMS was able to aquire over $240k$ GPU hours in the last year among several sites (see Figure \ref{fig-gpuhours}).

\section{Impact}
\label{sec-impact}

The advances of gravitational wave science and, multi-messenger astrophysics over the past four years have been absolutely contingent on the efficient and intelligent application of sophisticated, and computationally demanding, data analysis techniques.  Some of the greatest drivers of this demand come from efforts to characterize gravitational wave signals and estimate the parameters of the progenitor systems.   To meet the computational load these techniques demand, the gravitational wave community has significantly diversified its resource usage beyond dedicated sites to include more allocated and opportunistic resources.  The two parameter estimation pipelines which have spearheaded the usage of opportunistic resources are BayesWave~\cite{bayeswaves1} and RIFT~\cite{rift}.

These pipelines are routinely run in dedicated LIGO owned HTCondor computing clusters. In these clusters the run time environment (Operative System, libraries, etc) of a running job was highly controlled and uniform. Moreover the code was written to expect the input frame files to be in POSIX mounts at specific locations. Singularity provided the functionality to run a job inside a specific container, hence the solution to curate a distributed environment became to develop specific containers for each application. In its turn these containers would be distributed to all sites LIGO-owned or not using CVMFS \cite{singularity_cvmfs}. Moreover CVMFS is used in tandem with stashcache to provide a single POSIX-like mount at all computing sites to deliver the input data for the running jobs hence solving the data delivery problem as well.

BayesWave is an algorithm designed for robust signal classification and waveform reconstruction.    The ultimate goal here is to evaluate the respective Bayesian posterior probabilities that a given stretch of data contain a gravitational wave signal versus a transient “glitch” of terrestrial origin.  Posterior probability density functions on the parameters of a wavelet decomposition of the data are then used to reconstruct, or de-noise, the underlying signal. 

Under the hood, BayesWave utilizes a reversible-jump Markov chain Monte Carlo algorithm to explore a variable dimensional parameter space of gravitational wave signals, instrumental glitches and Gaussian noise.  Even when using this efficient, stochastic sampling algorithm, the confident BayesWave classification of a single, sub-second duration putative gravitational wave candidate as astrophysical or terrestrial in origin, and the reconstruction of the underlying waveform, can take up to 48 hours of computation on a single core.  

Furthermore, full characterization of a putative signal requires large scale Monte-Carlo simulations:  on the one hand, the statistical significance of a detection claim is ultimately determined by running the algorithm repeatedly on data which is believed to contain only noise; on the other hand, comparisons of waveform reconstructions with other analyses are quantified by running the BayesWave algorithm on thousands of simulations of the gravitational wave signal reported by those other analyses.

The properties of well-understood gravitational wave sources involving the coalescence of black holes and neutron stars are determined by comparing waveforms predicted by analytical or numerical models with data from the network of gravitational wave detectors.  The ultimate goal here is to generate a posterior probability density function from which we may select point estimates and credible intervals for the progenitor system’s parameters of interest, such as the mass, spin configuration, and so on.

While this is straightforward and well-posed in principle, a number of factors result in computationally costly analyses: a large and uncertain parameter space; evaluation of complex waveform models millions of times per source; and an often richly-structured, multi-modal likelihood function from which it is difficult to efficiently sample.   Even using simplified waveform models for binary neutron star mergers, these analyses can take hours and even weeks, depending on the extent of the parameter space and complexity of the model.  Sophisticated models for binary black hole mergers which include more exotic phenomenology may even take months to accurately determine the parameters of the system with confidence in the convergence of the results.  As with BayesWave, the cost is further compounded by the need to fully explore model systematics through large scale Monte Carlo simulations.

Rapid parameter Inference on gravitational wave sources via Iterative Fitting, i.e., RIFT, mitigates the costs inherent in sampling efficiency and waveform generation via a highly parallelizable grid-based algorithm.  Rather than sampling directly from the joint posterior probability distribution on all of the system’s parameters, RIFT constructs a grid over the intrinsic parameters (i.e., those which determine the system dynamics; typically the parameters of direct astrophysical interest) and employs Monte Carlo integration to marginalize over the extrinsic parameters (e.g., the spacetime coordinates for the event and its orientation with respect to Earth, which may be regarded as `nuisance’ parameters).  The marginalized likelihood of the intrinsic parameters is efficiently evaluated through generation of an initial cache of all possible model values at each grid point.  The grid of marginalized likelihood values then provides the seeds for Gaussian process interpolation to approximate the full, continuous likelihood function.  Samples from the target posterior distribution are then obtained via adaptive Monte Carlo techniques.    

The RIFT code has been significantly accelerated by leveraging GPUs.  After an initial CPU-bound calculation to evaluate inner products of the waveform models with the data, the matrix operations which yield the likelihood from these inner products and the marginalisation over extrinsic parameters are performed on a GPU.  A typical single-threaded CPU-bound job running on an Intel(R) Xeon(R) Silver 4116 completes with a wall time of about 7\,h 43\,m.  When the likelihood evaluation for the same job is performed on an Nvidia Quadro P2000, using the same CPU, the time to completion is just over 23\,m, a factor $\sim20\times$ improvement.  In terms of scientific analysis, a RIFT-based characterization of the GW170817 binary neutron star event requires 14 core-days, while a comparable analysis using more traditional sampling techniques requires 228 core-days.  The RIFT code is written in python, using CUPY to implement the CUDA-based GPU analysis.

Finally, to ensure convergence to a robust result, this procedure is applied iteratively, with the posterior samples from each stage providing a new grid for the subsequent stage, resulting in an adaptive grid refinement which accurately captures the shape of the likelihood function.  This algorithm lends itself naturally to a high-throughput computing approach, where each individual RIFT job independently explores a subset of the parameter space.

\section{Conclusions}
The functionalities of GlideinWMS and HTCondor have been sufficient to let the computing infrastructure help LIGO meet its scientific goals. PyCBC, BayesWaves and RIFT workflows have been successfully adapted to run in a Distributed High Throughput Computing model. This has led to an increase in the breadth and depth of the physics questions that can be answered by being able to consume hundreds of thousands of CPU and GPU hours in a worldwide distributed way. The near future still brings several short term challenges such as moving the infrastructure authentication from using X509 certificates to Scitokens and  placing submit hosts at several institutes outside the US and closer integration of the data delivery systems and the submission infrastructure.

\section{Conflict of Interest}

No conflict of interest exists:
We wish to confirm that there are no known conflicts of interest associated with this publication and there has been no significant financial support for this work that could have influenced its outcome.

\section*{Acknowledgements}

The authors would like to thank the funding agencies for this work, in particular the National Science Foundation through the following grants: OAC-1541349, OAC-1826967, OAC-1841530, MPS-1148698.



\bibliographystyle{elsarticle-num} 
\bibliography{main.bib}


\end{document}